\documentclass[12pt]{JHEP3}
\usepackage{amsmath}
\usepackage{amssymb}
\usepackage{bbold}

\newcommand{\eqn}[1]{(\ref{#1})}  
\newcommand{\adss}[2]{$\text{AdS}_{#1}\times S^{#2}$}

\makeatletter
\def\@clock{{\count0=\time
           \divide\count0 60
           \ifnum\count0<10 0\fi\the\count0
           \multiply\count0 -60 \advance\count0 \time
           :\ifnum\count0<10 0\fi \the\count0
         }}

\title{D-branes in a plane wave from covariant open strings}
\preprint{\hepth{0208038}\\CERN-TH/2002-177\\DAMTP-2002-98\\IC/2002/88}
\makeatother
\author{P. Bain\\DAMTP, University of Cambridge, \\ 
Centre for
Mathematical Sciences, Wilberforce Road, \\ 
Cambridge CB3 0WA, UK\\
{\tt E-mail: P.A.Bain@damtp.cam.ac.uk}} 
\author{K. Peeters\\
CERN, TH-division, \\
1211 Geneva 23,\\
Geneva, Switzerland\\
{\tt E-mail: Kasper.Peeters@cern.ch}}
\author{M. Zamaklar\\
The Abdus Salam ICTP, \\ 
Strada Costiera 11, \\ 
34014 Trieste,
ITALY\\
{\tt E-mail: mzama@ictp.trieste.it}}

\abstract{We derive boundary conditions for the covariant open string
corresponding to D-branes in an Hpp-wave, by requiring kappa symmetry
of its bulk action. Both half-supersymmetric and
quarter-supersymmetric branes are seen to arise in this way, and the
analysis furthermore agrees fully with the existing probe brane and
supergravity computations. We elaborate on the origin of dynamical and
kinematical supersymmetries from the covariant point of view. In
particular we focus on the D-string which only preserves half of the
dynamical supersymmetries and none of the kinematical ones. We discuss
its origin in \adss{5}{5} and its world-volume spectrum.}


\begin{document}
\section{Introduction}

Several recent papers have investigated D-branes in an Hpp-wave (a
homogeneous plane-wave solution of the type-IIB supergravity solution
with a constant five-form flux). A classification of their embeddings
and supersymmetries has been given by Skenderis and Taylor using a
probe brane approach~\cite{Skenderis:2002vf}, while equivalent results
have also been obtained by Bain~et~al.~by deriving supergravity
solutions of D-branes in Hpp-waves~\cite{Bain:2002nq}.  In the present
letter we would like to show that the analysis of boundary conditions
for the covariant open string leads to the same classification. Apart
from verifying the existing classification in an independent way, this
also sheds light on some recent questions concerning the quantum
analysis of these branes. In particular, we clarify the origin of
dynamical and kinematical supersymmetries from the covariant point of
view\footnote{Dynamical supersymmetries are those supersymmetries
which commute with the Hamiltonian, while kinematical supersymmetries
are those which do not.}. We restrict ourselves throughout to
so-called longitudinal branes, i.e.~branes for which the $x^+$ and
$x^-$ directions are part of the world-volume.

The Hpp-wave admits three classes of longitudinal branes without
world-volume fluxes. The first class consists of half-BPS branes,
which have straight embeddings in Rosen coordinates (they ``follow
geodesics of point-particles'') and have world-volume dimensions
$p=3,5,7$. The spectrum of these branes has been analysed by Dabholkar
and Parvizi~\cite{Dabholkar:2002zc} using the quadratic Green-Schwarz
action in the light-cone gauge as derived by
Metsaev~\cite{Metsaev:2001bj}. The analysis of Bergman et
al.~\cite{Bergman:2002hv} has shown these boundary conditions to be
consistent with open/closed string duality. The second class of branes
is obtained by moving these branes away from the origin, or formulated
more precisely, by embedding them along straight coordinate lines in
the Brinkman coordinate system. This breaks the dynamical
supersymmetries, so that only the kinematical supersymmetries remain.
They have been argued to be inconsistent with open/closed string
duality~\cite{Bergman:2002hv}. We will derive the corresponding open
string boundary conditions for these two classes of D-branes directly
from kappa symmetry requirements.

The third, and most interesting, class of branes consists of a single
one, namely a D-string which preserves half of the dynamical and none
of the kinematical supersymmetries. While it is at present not known
whether the presence of \emph{only} dynamical supersymmetries is
enough for quantum consistency (the corresponding boundary state has
not yet been constructed), there are several hints that suggest that
this object should be taken seriously. Firstly, it has been found as a
fully localised solution in supergravity~\cite{Bain:2002nq} (in
contrast, the seemingly inconsistent branes mentioned in the previous
paragraph only exist as smeared supergravity solutions). Secondly, we
will show that the D-string can be traced back to an unstable object
in the \adss{5}{5} geometry, and argued to become stable when the
Penrose limit is taken. An important consequence of the existence of
the D-string is that by a construction as in~\cite{Bergman:2002hv}, it
would imply the existence of a consistent D-instanton in the
Hpp-wave.\footnote{We thank Matthias Gaberdiel and Michael Green for
discussions about this issue.}

We will start in the next section with a derivation of the D-brane
boundary conditions from open string kappa symmetry. After
establishing which supersymmetries and kappa symmetries are preserved
by the various branes listed above in the covariant set-up, we
explicitly construct their realisation on the physical states in the
light-cone gauge in section~\ref{s:susies}. We show that there is a
one-to-one relation between the branes we find and those obtained with
other techniques, and moreover argue that all of them, including the
D-string, can be seen in the light-cone gauge. The AdS origin of the
D-string and its spectrum (in the Hpp-wave) is discussed in the last
section.  \vfill

\section{D-brane boundary conditions from kappa symmetry}
\subsection{Generalities}
\label{s:HppGS}

The Green-Schwarz action is invariant under local
\mbox{$\kappa$-symmetry}, which ensures that half of the fermionic
degrees of freedom can be gauged to zero such that the resulting
spectrum has the expected supersymmetry. For closed strings, the
constraint of \mbox{$\kappa$-symmetry} ensures that the background
fields satisfy their equations of motion. For open strings,
\mbox{$\kappa$-symmetry} transformations result in boundary terms,
which do not vanish without further constraints on the world-volume
fields at the boundary. For a Minkowski background, these boundary
terms have been examined by Lambert and West~\cite{Lambert:1999id} and
were shown to vanish when the standard boundary conditions for
half-BPS branes are imposed (a similar type of analysis for membranes
in eleven dimensions has been given by Ezawa et~al.~\cite{ezaw1} and
de Wit et~al.~\cite{deWit:1997fp}).  In the present section we will
extend the analysis to cover open Green-Schwarz strings in an Hpp-wave
background.

Most of this calculation is rather technical, so we present here only
the variations of the action and the resulting boundary conditions. At
up to fourth order in the fermions, these boundary terms arise from
variation of~\eqn{e:S1WZ} and~\eqn{e:firstMsquared}, labeled as
$S^1_{WZ}$ and $S^2_{WZ}$ respectively. Full details can be found in
the appendix.

\subsection{Branes ``at the origin''}

Let us first discuss branes at the origin of the coordinate
system.\footnote{Since the geometry is homogeneous, the coordinate
origin is equivalent to any other point in the space; branes sitting
at the origin can of course be located at any arbitrary other
location, but this is manifest only in Rosen coordinates. For
simplicity we will keep referring to these branes as ``branes at the
origin'' when we really mean ``branes which are flat in Rosen
coordinates''.}  The
terms in the kappa variation of the action (see equation~\eqn{e:S1WZ}
in the appendix) that survive in the flat space limit have appeared in
the literature before~\cite{Lambert:1999id}; they are given by
\begin{equation}
\label{e:flatboundaryterms}
\begin{aligned}
\delta_\kappa S^1_{WZ} \rightarrow 
 \int_{\partial\Sigma}
   \Big[ i  \big( \bar\theta_1 \Gamma_r \delta_\kappa\theta_1
            &- \bar\theta_2 \Gamma_r \delta_\kappa\theta_2 \big)\, {\rm
   d}X^\mu e_\mu{}^r
  \\
&  + \big(\bar\theta_1 \Gamma_r\delta_\kappa\theta_1\,\bar\theta_1 \Gamma^r {\rm d}\theta_1
      - \bar\theta_2 \Gamma_r\delta_\kappa\theta_2\,\bar\theta_2 \Gamma^r {\rm d}\theta_2
    \big)\Big]
  \, ,
\end{aligned}
\end{equation}
Here we used kappa symmetry to express the variations of the
bosons in terms of the variations of the fermions using~\eqn{e:defkappa}.
The terms above vanish by imposing the usual half-BPS boundary conditions,
\begin{equation}
\label{e:proj1}
\theta_1 = P \theta_2 \, ,\quad
\bar\theta_1 = \bar\theta_2 P\, (-)^{\frac{(p-1)(p-2)}{2}+p}
\end{equation}
with
\begin{equation}
\label{e:proj2}
P=\Gamma^{+- 1\cdots (p-1)}\, ,\quad  P^2 = (-)^{\frac{(p-1)(p-2)}{2}}\, .
\end{equation}
Since the background five-form breaks the SO(8) symmetry to
SO(4)$\times$SO(4), we will label these gluing matrices~$P$ by two numbers $n$
and $m$, denoting the number of gamma matrices with indices in the
first and second four transverse coordinates. The operators
$P^{(n,m)}$ satisfy the relations
\begin{equation}
P^{(n,m)} I    = I P^{(n,m)} \, (-)^{n}\,, \quad 
P^{(n,m)} \Gamma^r = \begin{cases} 
   \Gamma^r P^{(n,m)} (-)^{p+1} &\text{when $r\in D$,}\\[1ex]
   \Gamma^r P^{(n,m)} (-)^p     &\text{when $r\in N$,}
\end{cases}
\end{equation}
(where $N=\{+,-,1,\ldots,p-1\}$ and $D=\{ p, \ldots, d-2 \}$ and the
operator $I$ is defined in~\eqn{e:IandJdef}). One can then derive that
\begin{equation}
\label{e:standardflip}
(\bar\theta_1 \Gamma_r \delta_\kappa \theta_1) = 
  \begin{cases}
     -(\bar\theta_2 \Gamma_r \delta_\kappa \theta_2) &\text{when $r\in D$,}\\[1ex]
     +(\bar\theta_2 \Gamma_r \delta_\kappa \theta_2) &\text{when $r\in N$.}
  \end{cases}
\end{equation}
These relations clearly make~\eqn{e:flatboundaryterms} vanish.

The new terms with respect to flat space arise from the five-form
coupling in the Wess-Zumino term. We find that they lead to the
boundary terms
\begin{equation}
\label{e:deltaSWZQ}
\begin{aligned}
\delta_\kappa S^1_{WZ}\Big|_{\theta^2} &\rightarrow
  2\mu \int_{\partial\Sigma} 
   \Big[\begin{aligned}[t]
    {}& \big(\bar\theta_1 \Gamma_r \delta_\kappa\theta_1 +
         \bar\theta_2 \Gamma_r \delta_\kappa\theta_2 \big) \\
    {}& \quad\quad\times\big(\bar\theta_1 \Gamma^{[r} I \Gamma^+ \Gamma^{s]} \theta_2
        +\bar\theta_2 \Gamma^{[r} I \Gamma^+ \Gamma^{s]} \theta_1
    \big) \Big] {\rm d}X^\nu e_{\nu s}\, , \end{aligned}  \\[1ex]
\delta_\kappa S^1_{WZ}\Big|_{\theta^4} &\rightarrow 
  \frac{1}{2}\mu \int_{\partial\Sigma}
    \Big[ 
      \big( \bar\theta_1 \Gamma_r \delta_\kappa\theta_1\big)
      \big( \bar\theta_2 \Gamma^r I \Gamma^+ \Gamma^s \theta_1\big)
     + (1\leftrightarrow 2)
    \Big] {\rm d}X^\nu e_{\nu s}
\, .
\end{aligned}
\end{equation}
Note that these terms do \emph{not} cancel against each other using
only the half-BPS conditions (\ref{e:proj1}) and the flip
formula~\eqn{e:standardflip}. Some of the terms that arise by varying
$S^2_{WZ}$ in~\eqn{e:firstMsquared} are of a similar type, but they do
not come with the right coefficient to cancel against the variation of
$S^1_{WZ}$. Therefore, we need to cancel all of these contributions
separately by imposing appropriate additional boundary
conditions. Besides the relation~\eqn{e:standardflip} we now also need
the new flip relation
\begin{equation}
(\bar\theta_1 \Gamma^{[r} I \Gamma^+ \Gamma^{s]} \theta_2) = 
  \begin{cases}
    (-)^{\frac{(p-1)(p-2)}{2}+n+1}\,
    (\bar\theta_2 \Gamma^{[r} I \Gamma^+ \Gamma^{s]} \theta_1)
    &\text{when $r\in D$, $s\in N$,}\\[1ex]
    (-)^{\frac{(p-1)(p-2)}{2}+n\phantom{+1}}\,
    (\bar\theta_2 \Gamma^{[r} I \Gamma^+ \Gamma^{s]} \theta_1)
    &\text{when $r\in N$, $s\in N$.}\\[1ex]
  \end{cases}
\end{equation}
The variations in~\eqn{e:deltaSWZQ} are then seen to cancel
(separately for each line) when
\begin{equation}
\label{e:pncond}
\tfrac{1}{2}(p-1)(p-2) + n = \text{odd} \, .
\end{equation}
This condition also makes the boundary terms arising from
$\delta_\kappa S^2_{WZ}$ vanish.  We see that, unlike in Minkowski
space, kappa symmetry is preserved in the Hpp-wave only for particular
orientations of the D-branes in the wave (given by $n$ and $p=m+n+1$).
Equation~\eqn{e:pncond} holds true for the $(2,0)$, $(3,1)$ and
$(4,2)$ embeddings, which are the half-BPS branes with $p=3,5$ and $7$
respectively. These results are in agreement with the analysis of
probe branes~\cite{Skenderis:2002vf} and the light-cone open string
analysis of~\cite{Dabholkar:2002zc}.  In contrast, these boundary
terms do not vanish for the $(0,0)$ embedding, which is the D-string.

For this $(0,0)$ embedding, the first line in~\eqn{e:deltaSWZQ}
vanishes without further conditions because both $r$ and $s$ have to
be Neumann directions, i.e.~either plus or minus directions. However, canceling the second line
leads to one of the constraints
\begin{equation}
\label{e:d1string}
\Gamma^- \theta_{1,2} \Big|_{\partial\Sigma} = 0
\quad\text{or}\quad
\Gamma^+ \theta_{1,2}\Big|_{\partial\Sigma} = 0\, .
\end{equation}
The latter leads, as expected and as we will see below, to no further
problems when going to the light-cone gauge. The former condition is a
bit special, as it shows that the $(0,0)$ D-string can be described in
an alternative way. This condition seems to be sufficient to make all
boundary terms at higher orders in theta vanish as well (see the
appendix). It is, however, incompatible with the standard light-cone
gauge.

With the above conditions, one can verify that the boundary terms
arising from $S^2_{WZ}$ vanish as well; we will not discuss these in
detail here.

\subsection{Branes ``outside the origin''}

Let us now discuss the spin-connection dependent terms in the
variation of the action. In contrast to the terms discussed above,
these terms depend on the location of the end-point of the string in
the target space. As we will see they automatically vanish for all
D-branes sitting at the origin of the coordinate system. However, for
branes outside the origin (and with flat embedding in Brinkman
coordinates), their cancellation leads to additional boundary
conditions. We find the following boundary terms:
\begin{equation}
\begin{aligned}
\delta_\kappa S^1_{WZ} \Big|_{\theta^2} &\rightarrow \int_{\partial\Sigma}
   \Big[\begin{aligned}[t]
   {}&  \big(\bar\theta_1\Gamma^n \delta_\kappa\theta_1 
          + \bar\theta_2\Gamma^n\delta_\kappa\theta_2\big)\\
   {}&  \quad\quad\times\big(\bar\theta_1 \Gamma_{[n}{}^{rs} \omega_{m]rs} \theta_1
          -\bar\theta_2 \Gamma_{[n}{}^{rs} \omega_{m]rs} \theta_2\big)
   \Big]{\rm d}X^\mu e_\mu{}^m \, ,\end{aligned} \\[1ex]
\delta_\kappa S^1_{WZ} \Big|_{\theta^4} &\rightarrow
     \frac{1}{4}\int_{\partial\Sigma}
   \Big[
     (\bar\theta_2\Gamma^n\delta_\kappa\theta_2)
     (\bar\theta_1\Gamma_n{}^{rs} \omega_{mrs} \theta_1) 
      - (1\leftrightarrow 2)
   \Big] {\rm d}X^\mu e_\mu{}^m \, .
\end{aligned}
\end{equation}
For the first line, we only have to consider the case where the $m$ or
$n$ index on the gamma matrix in the second factor is not $+$ or $-$ (a
lower minus sign on the gamma matrix is excluded because
$\omega_{++i}$ is the only non-vanishing component of the spin
connection, see~\eqn{e:spinconnPP};  a lower plus sign on the gamma
matrix would mean that both $n$ and $m$ are plus, and
anti-symmetrisation then sets everything to zero).  Using the exchange
property~\eqn{e:standardflip}, the boundary terms then reduce to
\begin{equation}
\label{e:deltaSWZom}
\begin{aligned}
\delta_\kappa S^1_{WZ} \rightarrow{} & \mu\int_{\partial\Sigma}
   \Big[\big(\bar\theta_1\Gamma^+ \delta_\kappa\theta_1 \big) {\rm d}X^n      
       -\big(\bar\theta_1\Gamma^n \delta_\kappa\theta_1 \big) {\rm d}X^+
   \Big] \big( \bar\theta_1 \Gamma_{n}{}^{+s'} \theta_1-\bar\theta_2 \Gamma_{n}{}^{+s'} \theta_2\big)
  \partial_{s'} S \\[1ex]
& \quad\quad- \frac{1}{4}\Big[ \big(\bar\theta_1 \Gamma^n \delta_\kappa\theta_1\big) 
      \big( \bar\theta_2 \Gamma_{n}{}^{+s'} \theta_2\big) -
   (1\leftrightarrow 2) \Big] {\rm d}X^+ \partial_{s'} S\, ,
\end{aligned}
\end{equation}
where prime on the index $s$ indicates that it no longer takes the
values $+$ or $-$.  We now need the exchange property
\begin{equation}
(\bar\theta_1\Gamma_{mn-}\theta_1) = \begin{cases}
+(\bar\theta_2\Gamma_{mn-}\theta_2)  & \text{when $(m,n) \in (D,D)$ or $(N,N)$,} \\[1ex]
-(\bar\theta_2\Gamma_{mn-}\theta_2)  & \text{when $(m,n) \in (N,D)$ or $(D,N)$.}
\end{cases}
\end{equation}
This can be used to show that the above boundary
terms~\eqn{e:deltaSWZom} vanish only when the derivative on~$S$ is in
a Neumann direction. When the derivative is in the Dirichlet
direction, the above boundary terms are non-zero outside the
origin. In this case we need to impose
\begin{equation}
\label{e:otherquarter}
\Gamma^+ \theta_{1,2}\Big|_{\partial\Sigma} = 0\, 
\end{equation}
in order to make the boundary terms vanish. As in the previous
section, these conditions also make the boundary terms arising from
$S^2_{WZ}$ vanish. We will see the implications of these conditions
for the remaining supersymmetry on physical states in the next
section.

\section{Kinematical and dynamical supersymmetries}
\label{s:susies}
\subsection{Symmetries, constraints and gauge fixing}

Having derived D-brane boundary conditions for the open string
variables, we now want to analyse their consequences in terms of the
remaining supersymmetries on physical states. In order to make contact
with results obtained using probe branes or supergravity solutions, we
first need to understand the relation between the supersymmetry
parameters appearing in the transformation rules of the \emph{closed}
string action and those appearing in the supergravity
transformations. After that, we need to investigate the \emph{open}
string and determine which subset of these symmetries preserves the
boundary conditions. Finally, we fix the kappa gauge freedom and
obtain the action of the global symmetries on the physical states of
the open string with D-brane boundary conditions.

Let us start by discussing the global supersymmetry invariance of the
covariant action, i.e.~before fixing a \mbox{$\kappa$-symmetry} gauge
(see also~\cite{deWit:1998tk}). In general the action is \emph{by
construction} invariant under the transformations\footnote{These
transformation rules are only correct to lowest order in theta. We
are suppressing higher-order theta contributions here as they later
become irrelevant anyway when we go to the light-cone gauge.}
\begin{equation}
\delta \theta_{1,2} = \epsilon_{1,2} \,,\quad
\delta X^\mu = i\bar\theta_1\Gamma^\mu \epsilon_1 +
i\bar\theta_2\Gamma^\mu \epsilon_2
\, ,
\quad \epsilon = \epsilon_1 +
i\epsilon_2\, ,
\end{equation}
for an \emph{arbitrary} (not necessarily constant) target space spinor
$\epsilon$ evaluated at the world-volume, together with a
transformation of all the background fields using the supergravity
transformation rules with parameter $-\epsilon$. This is simply a reflection of the fact that superspace
diffeomorphisms are equivalent to component supersymmetry
transformations, and does not really constitute a non-trivial
symmetry. From this observation it follows, however, that for the
special situation in which~$\epsilon$ is a Killing spinor (evaluated
at the world-sheet) and the background is bosonic, the string action
is invariant under
\begin{equation}
\delta \theta_{1,2} = \epsilon_{1,2}^{\text{Killing}} \, ,\quad
\delta X^\mu = i\bar\theta_1\Gamma^\mu \epsilon^{\text{Killing}}_1 +
i\bar\theta_2\Gamma^\mu \epsilon^{\text{Killing}}_2
\end{equation}
by itself. This follows because the transformations of the background
fields become trivial in this case. In other words, the action is
invariant under \emph{global} supersymmetries generated as shifts,
with parameters which are identical to the target-space Killing
spinors. We will call these supersymmetries \emph{shift
symmetries}. We will in general suppress the label ``Killing'' on the
associated $\epsilon$ parameters. Note that the shift parameter
$\epsilon$ has a \emph{fixed}, but generically \emph{non-constant}
dependence on the string world-sheet coordinates.

Not all of the shift symmetries survive in the open string, as some of
them may be incompatible with the boundary conditions on the
$\theta_{1,2}$ fields. The remaining global
supersymmetries are easily seen when acting on physical fields 
(i.e.~in  the light-cone gauge, in which the  kappa symmetry
is eliminated).
Let us recall how to go to the
semi-light-cone gauge given by the condition
$\Gamma^+\theta_{1,2}=0$.\footnote{It is rather simple to show, as we
will do below, that this gauge is compatible with the boundary
conditions. Note, however, that the D-string can be obtained with an
alternative set of boundary conditions, namely
$\Gamma^-\theta_{1,2}\Big|_{\partial\Sigma}=0$;
see~\eqn{e:d1string}. These would pose problems for the standard
semi-light-cone gauge.} For
this we need to be more specific about the form of kappa
transformations. These act on fermions according to
\begin{equation}
\delta_\kappa\theta_{1,2} = \Gamma_r \Pi_{j}^r\, \kappa_{1,2}^j\, ,
\end{equation}
where $\Pi_i{}^r = \partial_i X^M E_M{}^r$.  The two kappa parameters
are self-dual and anti self-dual respectively,
\begin{equation}
\kappa_1^\tau = -\kappa_1^\sigma\, ,\quad
\kappa_2^\tau = \kappa_2^\sigma\, .
\end{equation}
One then easily obtains the kappa symmetry transformation necessary to
bring the action into the $\Gamma^+\theta_{1,2}$ gauge:
\begin{equation}
\label{e:kappatolc}
\kappa_{1,2}^\tau = -
\frac{1}{2}\frac{1}{\Pi_\tau^+ \mp \Pi_\sigma^+} \Gamma^+ \theta_{1,2}
\, .
\end{equation}
In the presence of a D-brane boundary condition 
\begin{equation}
\label{e:guggenheim}
\Gamma^+ \theta_{1,2}\Big|_{\partial\Sigma} = 0\,,
\end{equation}
the story is slightly changed, as this constraint fully removes kappa
symmetry at the boundary. However, even with this ``reduced'' kappa
symmetry (with $\kappa \rightarrow 0 |_{\partial \Sigma}$) we can
still globally go to semi-light-cone gauge, since all fermions on the
world sheet are also constrained by~\eqn{e:guggenheim}. In contrast,
the light cone gauge cannot be reached for the first D-string
condition in~\eqn{e:d1string}, as it would lead to discontinuous
fermionic fields on the world-volume.

Once in the $\Gamma^+ \theta_{1,2}=0$ gauge, half of the shift
supersymmetries ($\Gamma^+ \Gamma^- \epsilon$) are such that they keep
the system in this gauge. The other half of the shift symmetries
($\Gamma^- \Gamma^+ \epsilon$) moves the system out of the light-cone
gauge.  However, it is always possible to perform a compensating kappa
transformation, such that the gauge condition is restored. The
remaining global symmetries are then given by
\begin{equation}
\label{e:globalsusies}
\begin{aligned}
\delta\theta_{1,2} &= \frac{1}{2}\Gamma^+\Gamma^- \epsilon_{1,2} + 
\frac{1}{2} \Gamma^-\Gamma^+ \epsilon_{1,2} - 
\frac{1}{2}\frac{1}{\Pi_\tau^+\mp\Pi_\sigma^+} \Gamma_r (\Pi_\tau^r
\mp \Pi_\sigma^r) \Gamma^+\epsilon_{1,2}\\[1ex]
&= \frac{1}{2}\Gamma^+\Gamma^- \epsilon_{1,2} -
\frac{1}{2}\frac{1}{\Pi_\tau^+\mp\Pi_\sigma^+} \Gamma_{r'} (\Pi_\tau^{r'}
\mp \Pi_\sigma^{r'}) \Gamma^+\epsilon_{1,2}\, ,
\end{aligned}
\end{equation}
where primes indicate summation over transverse directions only.

All of these expressions are rather ugly unless a further bosonic
light-cone choice is made. When $X^+ = \tau$ the above results
simplify because ${\Pi_\tau^+\mp\Pi_\sigma^+}=1$. The remaining
covariant momenta $\Pi_i^{r'}$ also simplify to $\partial_i X^{r'}$ in
the Hpp-wave background.

\subsection{Supersymmetries in the light-cone gauge}

Let us now apply the general logic of the previous subsection to the
D-brane boundary conditions found before. The Killing spinors of the
Hpp-wave were constructed by Blau et~al.~\cite{Blau:2001ne} and are
in our conventions given by
\begin{multline}
\epsilon = \Big(1 + \sum_{m=1,2,3,4} \frac{i}{2} \mu X^m \Gamma^+ \Gamma_m I +
\sum_{m=5,6,7,8} \frac{i}{2} \mu X^m \Gamma^+
\Gamma_m J \Big)\\[1ex]
\times \Big( \cos^2(\tfrac{1}{2}\mu X^+) \mathbb{1} - \sin(\tfrac{1}{2}\mu X^+)^2 IJ 
      - i \sin(\tfrac{1}{2}\mu X^+) \cos(\tfrac{1}{2}\mu X^+) (I+J) \Big)
(\lambda + i \eta) \, .
\end{multline}
The spinors $\lambda$ and $\eta$ are constant. It is convenient to
decompose this expression using
\begin{equation}
\lambda = \tfrac{1}{2}\Gamma^+\Gamma^- \lambda +
\tfrac{1}{2}\Gamma^-\Gamma^+\lambda
:= \lambda^{(+)} + \lambda^{(-)}\, ,
\end{equation}
and similarly for the $\eta$ spinor. As the spinors have positive
chirality, one deduces
\begin{equation}
I J \big(\lambda^{(\pm)} + i\eta^{(\pm)} \big) = \pm \big(\lambda^{(\pm)} + i\eta^{(\pm)}
\big)\, ,\quad
I \big(\lambda^{(\pm)} + i\eta^{(\pm)} \big) = \pm J \big(\lambda^{(\pm)} + i\eta^{(\pm)}
\big)\, .
\end{equation}
We then obtain a decomposition of the Killing spinor into
$X^+$~dependent and $X^+$~independent parts,
\begin{equation}
\label{e:epsdecompose}
\begin{aligned}
\epsilon = \Big[ \cos^2(\tfrac{1}{2}\mu X^+) - \sin^2(\tfrac{1}{2}\mu X^+) - 2 i \sin(\tfrac{1}{2}\mu
X^+)\cos(\tfrac{1}{2}\mu X^+) I \Big] \big(\lambda^{(+)} + i\eta^{(+)}\big) & \\[1ex]
+ \Big[ 1 + \frac{\mu}{2}\Big(\sum_{m=1,2,3,4} - \sum_{m=5,6,7,8} \Big) 
i X^m \Gamma^+ \Gamma_m I \Big] \big(\lambda^{(-)} + i\eta^{(-)} \big) & \, .
\end{aligned}
\end{equation}
These two lines correspond to the ``kinematical'' and ``dynamical''
part of the Killing spinor. Note, however, that this decomposition
does \emph{not} correspond to the decomposition $\epsilon =
\frac{1}{2}\Gamma^+\Gamma^-\epsilon + \frac{1}{2}\Gamma^-\Gamma^+\epsilon$.

Using expression~\eqn{e:globalsusies} we can now immediately write
down the \emph{global} supersymmetry invariances of the \emph{closed}
Green-Schwarz action in the Hpp-wave\footnote{Similar global
transformation rules where given for the $d=11$~supermembrane by
Sugiyama and Yoshida~\cite{Sugiyama:2002rs}; their derivation, however,
does not start from the covariant action.}:
\begin{align}
\label{e:deltatheta1}
\delta\theta_1 &=
 \tfrac{1}{2}\Gamma^+\Gamma^- \bigg[ \left(\cos^2(\tfrac{1}{2}\mu X^+)
- \sin^2(\tfrac{1}{2}\mu X^+)\right) \lambda^{(+)} + 2\sin(\tfrac{1}{2}\mu X^+)\cos(\tfrac{1}{2}\mu X^+) I
\eta^{(+)} \bigg]\nonumber\\[1ex]
{}& - \tfrac{1}{2} \Gamma_{r'} (\partial_\tau-\partial_\sigma) X^{r'}
\Gamma^+ \lambda^{(-)}\nonumber\\[1ex]
{}& - \tfrac{1}{4} \Gamma^+\Gamma^- \bigg[ \mu \Big( \sum_{m=1,2,3,4} -
\sum_{m=5,6,7,8} \Big) X^m \Gamma^+ \Gamma_m I \eta^{(-)} \bigg]\, ,
\\[1ex]
\label{e:deltatheta2}
\delta\theta_2 &= 
 \tfrac{1}{2}\Gamma^+\Gamma^- \bigg[ \left(\cos^2(\tfrac{1}{2}\mu X^+)
- \sin^2(\tfrac{1}{2}\mu X^+)\right) \eta^{(+)} - 2\sin(\tfrac{1}{2}\mu X^+)\cos(\tfrac{1}{2}\mu X^+) I
\lambda^{(+)} \bigg]\nonumber\\[1ex]
{}& - \tfrac{1}{2} \Gamma_{r'} (\partial_\tau+\partial_\sigma) X^{r'}
\Gamma^+ \eta^{(-)}\nonumber\\[1ex]
{}& + \tfrac{1}{4} \Gamma^+\Gamma^- \bigg[ \mu \Big( \sum_{m=1,2,3,4} -
\sum_{m=5,6,7,8} \Big) X^m \Gamma^+ \Gamma_m I \lambda^{(-)} \bigg]\, .
\end{align}
These are accompanied by a transformation of the transverse bosons,
which get a contribution both from the shift symmetry as well as the
compensating kappa transformation,
\begin{equation}
\delta X^{r'} = 2i\,\bar\theta_1 \Gamma^{r'}\lambda^{(-)}
                +2i\,\bar\theta_2 \Gamma^{r'}\eta^{(-)}\, .
\end{equation}
Instead of relying on the covariant arguments given so far, one can of
course also verify directly that the action of a \emph{closed} string
\begin{multline}
\label{e:lcaction}
S = \int\!{\rm d}^2 \sigma \Big[
- \frac{1}{2}\partial_+ X^{r'} \partial_- X_{r'} 
- \frac{1}{2}\mu^2 X^{r'} X_{r'}\\
+ i\, \bar\theta_1\Gamma^{-}\partial_+ \theta_1
+ i\, \bar\theta_2\Gamma^{-}\partial_- \theta_2
- 2i\mu\, \bar\theta_1 \Gamma^{-} \Pi \theta_2
\Big]\, ,
\end{multline}
(with $\partial_\pm = \partial_\tau\pm\partial_\sigma$) is indeed
invariant under the symmetries given above. 

As we discussed before, not all of these symmetries of the closed
string survive for an \emph{open} string, as the boundary conditions
can be such that some of the global
supersymmetries~\eqn{e:deltatheta1} and~\eqn{e:deltatheta2} do not
survive. Let us, as an example, discuss the D-string boundary
conditions. By imposing $\delta\theta_1 = \Gamma^{+-} \delta\theta_2$
at the boundary, we find from~\eqn{e:deltatheta1}
and~\eqn{e:deltatheta2} that
\begin{equation}
\label{e:constraint_D1}
\lambda = \Gamma^{+-}\eta\, ,\quad
\lambda^{(+)} = \eta^{(+)} = 0\, ,
\end{equation}
which together imply $\lambda^{(-)}=-\eta^{(-)}$. When this relation
holds, one indeed verifies that the action~\eqn{e:lcaction} for an
open string is invariant. This condition is precisely what has been
obtained from probe branes as well. In the next subsection we will
make this correspondence more precise for general branes.

\subsection{Comparison with probe brane and supergravity results} 

A nice consequence of the above analysis is that it makes it very
simple to show that the global supersymmetries preserved by the open
string with various D-brane boundary conditions indeed match the probe
brane results of Skenderis and Taylor~\cite{Skenderis:2002vf}. The
condition for kappa symmetry of the probe brane embedding is written
as
\begin{equation}
\epsilon = \gamma \epsilon \,,\quad \text{with}\quad
\gamma = \gamma^{+-mn} (*)^{\frac{p+1}{2}} (-i)\, ,
\end{equation}
where $\gamma$ is the kappa symmetry projector and $m$ and $n$
symbolically denote the number of indices in the first and second four
transverse directions. This condition can be rewritten as
\begin{equation}
\label{e:fromprobe}
\epsilon_1+i\epsilon_2 = \Gamma^{+-m n} \big(\epsilon_2 - (-)^{\frac{p+1}{2}} i\epsilon_1\big)\quad
\Rightarrow \quad \epsilon_1 = \Gamma^{+-mn} \epsilon_2\, ,
\end{equation} 
(the real and imaginary part of the equation are equivalent because of
the relation \mbox{$(\Gamma^{+-mn})^2=(-)^{\frac{1}{2} p(p+1) -1}$}
which for odd $p$ equals \mbox{$-(-)^{\frac{p+1}{2}}$}). Requiring a
match between linearly independent terms of this equation is then
identical to the conditions obtained from $\delta\theta_1 =
\Gamma^{+-mn}\delta\theta_2$, thereby completing the proof of the
equivalence between probe brane and open string results.

A match is also found by comparing with the supergravity solutions of
Bain et~al.~\cite{Bain:2002nq}. Here the comparison is necessarily
less systematic, as the Killing spinors of the brane-in-wave
backgrounds are not simply obtained from the Killing spinors of the
Hpp-wave. The various branes have to be addressed case-by-case. For
e.g.~the D-string solution, we see from equation~(3.18)
of~\cite{Bain:2002nq} that it requires
$\gamma^+\gamma^-\hat\epsilon=0$, which in our notation corresponds
to~\eqn{e:constraint_D1}.

\section{The ``quarter BPS'' D-string}
\subsection{\adss{5}{5} origin}

Boundary conditions for open strings which are consistent at
tree level do not necessarily have to be consistent for higher genus
open string surfaces. This was demonstrated by Bergman
et~al.~\cite{Bergman:2002hv} for D$p$-branes with $p>1$ located
outside the origin of the Brinkman coordinates. Tree level boundary
conditions used at one-loop in the open string genus expansion seem to
be incompatible with the open/closed string duality.
Having established consistent tree level open string boundary
conditions for the quarter-supersymmetric D-string, we therefore now
have to investigate whether this solution makes sense in the full
theory.\footnote{Calling the D-string ``quarter-BPS'' is perhaps not
quite correct, as BPS conditions can only be derived from supercharges
that square to the Hamiltonian. Since the kinematical supercharges do
not have this property, the BPS fraction strictly speaking only refers
to the number of unbroken dynamical supersymmetries. In this sense the
D-string could be called ``half-BPS'', but to avoid confusion we will
keep referring to all branes by the total number of unbroken
supersymmetries.}

One way of attacking this question is to follow an analysis similar
to~\cite{Bergman:2002hv}. Another, less direct argument is to try to
trace back the D-string to the AdS and D3-brane geometries.  One
expects that all branes which are consistent in the D3-brane geometry
should also be consistent in the AdS geometry, and these should in
turn lead to consistent D-branes in the Hpp-wave. Hence, if one can
prove that the particular D-string in the D3-brane or AdS background
which leads to the D-string in the Hpp-wave geometry (after the
sequence of near-horizon and Penrose limits) is consistent in the
initial space, then the D-string should also be a consistent solution
in the Hpp-wave.

Although the D-string in the AdS space can not be quantised and
analysed directly, the fact that there should be a dual gauge
description might be useful in proving its existence. Unfortunately
the dual gauge description of the string is at this moment not yet
fully under control~\cite{kmgauge}. However, we would like to mention several
properties of this D-string.

It is obvious, from the way the Penrose limit is taken, that the
D-string which wraps the big circle of the $S^5$ through which the
limiting null geodesic passes leads to the D-string at the origin of
the Hpp-wave space.  This D-string is unstable against small
perturbations, and will collapse to a point (i.e.~a minimal $S^1$ on
$S^5$) as we will now discuss.

Since the open string/CFT description of this D-string is lacking,
we can use the effective action approach to study its (in)stability by
looking at the spectrum of small quadratic fluctuations around the
 static position. Therefore, we start with the \adss{5}{5}
space 
written in global coordinates
\begin{equation}
{\rm d}s^2 = R^2 (
-{\rm d}t^2 \cosh^2\rho + {\rm d} \rho^2 + \sinh^2\rho\, {\rm d} \Omega_3^2 +
{\rm d} \psi^2 \cos^2 \theta + {\rm d}\theta^2 + \sin^2\theta\,
{\rm d} {\Omega_3'}^2 )
\end{equation}
and consider the D-string wrapping the equator of $S^5$. Choosing
the static gauge $t=\tau$ and $\psi=\sigma$ with all the other
coordinates set to zero, it is easy to see that this embedding is a
solution of the Dirac-Born-Infeld equations of motion
\begin{equation}
\label{e:dbi_eq}
\partial_i \left(\sqrt{-{{\rm det} \hat{g}}}\, \hat{g}^{ij} \partial_i
X^\nu g_{\mu\nu}\right) - \frac{1}{2} 
\sqrt{-{{\rm det} \hat{g}}}\, \hat{g}^{ij}\partial_i X^\nu \partial_j
X^\rho \partial_\mu g_{\nu\rho} = 0\, ,
\end{equation} 
where $\hat{g}$ is the induced metric on the D-string world-sheet. 

To investigate the issue of stability, it is enough to look at the
fluctuations in the angular direction $\theta$. We will also consider
the fluctuations in the AdS space direction $\rho$ to show that they
do not contain unstable modes. We introduce arbitrary functions
$\delta\theta$ and $\delta\rho$ of the world-volume coordinates and
expand the DBI action to quadratic order in these fluctuations
\begin{equation}
S_{\text{quadr.}} \sim \int\!{\rm d}^2\sigma\, 
\left(- \delta\dot{\rho}^2 + \delta{\rho'}^2 + \delta\rho^2 -
\delta\dot{\theta}^2 + \delta{\theta'}^2 - \delta \theta^2\right) \,  . 
\end{equation}
From this expression we conclude that the zero mode of $\delta \theta$
is indeed tachyonic, signaling the instability of the solution under small
perturbations in this direction.

Actually, it is possible to describe in detail the behaviour of this
perturbed D-string by searching for a time dependent solution $\theta(t)$
of the equations of motion~\eqn{e:dbi_eq}. 
For $\mu=\psi$, this equation is satisfied as in the unperturbed case
while for $\mu=t$, it leads to 
\begin{equation}
\label{e:theta_eq}
\frac{d}{dt}\left(\frac{\cos\theta}{\sqrt{1-\dot{\theta}^2}}\right) = 0 
\end{equation} 
and it is easy to check that the equation~\eqn{e:dbi_eq} for
$\mu=\theta$ is a consequence of~\eqn{e:theta_eq}. Integrating one
time~\eqn{e:theta_eq} gives the first order differential equation 
\begin{equation}
\label{e:theta_eq2}
\frac{{\rm d}{\theta}}{\sqrt{1 - c^2 \cos^2\theta}} =  {\rm d}t\, ,
\end{equation}
where $c$ is a constant which is determined in terms of the
initial position $\theta_0$ and velocity $v_0\equiv\dot\theta_0$~: 
$c = \sqrt{1-v_0^2}/\cos\theta_0$.  The
equation~\eqn{e:theta_eq2} can be 
integrated by using a function known as the incomplete elliptic
integral in the mathematical literature and denoted $F(x \vert c)$~:
\begin{equation}
\label{e:theta_sol}
F(\theta \vert c)- F(\theta_0 \vert c)=  t. 
\end{equation}
which defines implicitly the evolution of the D-string on the
5-sphere. When $c=1$, which encompasses the case of the static D-string
sitting at the equator, we can give a more explicit expression
\begin{equation}
\label{e:theta_sol2}
\theta(t) = \arcsin\left(\frac{2 a e^t}{1+a^2 e^{2t}}\right) \qquad {\rm with}
\qquad a \equiv \sqrt{\frac{1-\cos\theta_0}{1+\cos\theta_0}}\,.
\end{equation}
From this relation, one can write down the induced metric on the
world-volume of the D-string
\begin{equation}
\begin{aligned}
{\rm d}s^2_{\rm D1} &= R^2(- {\rm d}t^2 + {\rm d}\theta^2 
                      + \cos^2 \theta {\rm d}\psi^2) 
= R^2\,\left(\frac{1-a^2 e^{2t}}{1+a^2 e^{2t}}\right)^2\, 
(- {\rm d}t^2 + {\rm d}\psi^2) 
\end{aligned}
\end{equation}
and compute, as a function of its initial position, the time it takes
for the D-string to collapse to one of the poles of $S^5$~: $t_s = -
\log a = {\rm arctanh}\left(\cos\theta_0\right)$. As expected, this time
goes to infinity when the D-string 
is initially infinitesimally close to the equator ($\theta_0 \rightarrow 0$). 

Let us now consider the Penrose limit of this shrinking D-string. The
time $t$ and  active angular coordinates $\psi$ and $\theta$ are
rescaled as  
\begin{equation}
\begin{aligned}
t = x^+ + \frac{x^-}{R^2}\, , \qquad 
\psi = x^+ - \frac{x^-}{R^2} \qquad {\rm and} \qquad 
\theta = \frac{y}{R} \qquad {\rm for} \qquad R \rightarrow \infty.
\end{aligned}
\end{equation}
Therefore, we observe that, in order to survive under the Penrose
limit, the D-string defined by the equation~\eqn{e:theta_sol2} must
sit initially at an angle $\theta_0$ such that $a \sim 1/R$. In other
words, it must be at a distance $\theta_0$ of order $1/R$ from the
equator.  This analysis can be repeated in the general case $c\neq 1$
and one can see that the D-strings which survive must again be
infinitesimally close (in initial position and velocity) to the static
configuration; in more quantitative words, one should also have $c-1
\sim 1/R^2$.  The Penrose limit of such D-strings correspond to stable
solutions of the DBI action in the plane wave background. In
particular, the tachyonic mode along the direction $\theta$ is washed
out under this process. This can be seen by evaluating the action on
the Penrose limit of the solution; in contrast to the action of the
shrinking D-string in the AdS background, the action is now constant.
Hence, through the Penrose limit, the \emph{unstable} (static or
moving) D-string on $S^5$ gets mapped into a \emph{susy} and hence
stable (static or moving) configuration in the Hpp-wave geometry.

Finally, let us conclude this section by observing that the
D-string in the AdS space which is wrapping the equator of the
5-sphere originates 
from an unstable circular D-string in the geometry of the D3-branes,
which lies in the two-plane that intersects the D3-brane over a point,
times the time. This is obvious from the isometries which are preserved by
these two configurations. Note, however, that while the circular
string in the D3-brane geometry is always a \emph{non-static}
configuration, this is not true any longer in the near horizon
geometry, which admits a static configuration.

\bigskip
\subsection{World-volume spectrum}

In order to find the world-volume spectrum we have to determine the
mode expansion of the bosonic and fermionic fields on the open
string. The equations of motion obtained from the light-cone
action~\eqn{e:lcaction} are
\begin{equation}
\begin{aligned}
\partial_+\partial_- X^r + \mu^2 X^r &= 0\, ,\\[1ex]
\partial_+ \theta^1 - \mu I \theta^2 &= 0\, ,\\[1ex]
\partial_- \theta^2 + \mu I \theta^1 &= 0\, .
\end{aligned}
\end{equation}
The mode expansions for the closed string can be found
in~\cite{Metsaev:2001bj} and these have been used to construct the
mode expansions subject to half-supersymmetric boundary conditions,
see~\cite{Dabholkar:2002zc}. For boundary conditions that remove all
of the kinematical supersymmetries the story is slightly different, as
we will now show. We will for simplicity only consider strings
starting and ending on a D-string, i.e.~with boundary conditions
\begin{equation}
\theta^1 = \Gamma^{+-}\theta^2 \Big|_{\sigma=0,\pi}\, .
\end{equation}
The $\tau$-independent parts of the fermionic mode expansion are easily
found. Using the ansatz
\begin{equation}
\label{e:theta_mode_expansion}
\begin{aligned}
\theta^1 &= (1+\Omega I)\, \theta^+ e^{\mu \sigma} + 
           (1-\Omega I)\, \theta^- e^{-\mu \sigma}
+ \sum_{n=-\infty}^{+\infty} 
c_n\Big(\theta^1_n        e^{-i\omega_n \tau - i n \sigma} 
                   +\tilde\theta^1_n  e^{-i\omega_n \tau + i n \sigma}\Big)
                   \,,\\[1ex]
\theta^2 &= (\Omega+ I)\, \theta^+ e^{\mu \sigma} + 
           (\Omega- I)\, \theta^- e^{-\mu \sigma}
+ \sum_{n=-\infty}^{+\infty} 
c_n\Big(\theta^2_n        e^{-i\omega_n \tau - i n \sigma} 
                   +\tilde\theta^2_n  e^{-i\omega_n \tau + i n \sigma}\Big)
                   \,,
\end{aligned}
\end{equation}
(where $\Omega=\Gamma^{+-}$) we find that, for the D-string boundary
conditions, the Fourier coefficients $\theta^1_n$, $\tilde\theta^1_n$,
$\theta^2_n$ and $\tilde\theta^2_n$ can all be expressed in terms of a
single mode $\theta_n$ as
\begin{equation}
\begin{aligned}
\theta^1_n &=
\left[
n - \omega_n + i {\mu} \Omega I
\right] \theta_n\, ,\\[1ex]
\tilde\theta^1_n &=
\left[
n + \omega_n - i {\mu} \Omega I
\right] \theta_n\, ,\\[1ex]
\theta^2_n &= 
\left[
n + \omega_n + i {\mu} \Omega I
\right] \Omega\theta_n\, ,\\[1ex]
\tilde\theta^2_n &= 
\left[
n - \omega_n -  i {\mu} \Omega I
\right] \Omega\theta_n\, ,
\end{aligned}
\end{equation}
where $\omega_n = {\rm sgn}(n)\sqrt{\mu^2 + n^2}$ and, for
convenience, we have introduced the factors $c_n =
1/\sqrt{1+{(\omega_n-n)^2}/{\mu^2}}$.  Note in particular that for
$n=0$ the above implies \mbox{$\theta_0^1 + \tilde\theta^1_0=0$} and
\mbox{$\theta_0^2 + \tilde\theta^2_0 =0$} which eliminates the zero
modes from the expansions~\eqn{e:theta_mode_expansion} (consistent
with the fact that these D-branes do not have any kinematical
supersymmetries). Finally, reality of the coordinates
implies that $\theta_{-n} = -(\theta_n)^\dag$.

The bosonic mode expansions have been given in \cite{Billo:2002ff},
\begin{equation}
\label{e:x_mode_expansion}
\begin{aligned}
X^r = 
\frac{x^r_0 \sinh \mu(\pi-\sigma) + x^r_1 \sinh \mu \sigma}
{\sinh\mu \pi}
+ i\sum_{n\neq 0} \frac{1}{\sqrt{2\vert\omega_n\vert}} \, 
a^r_n e^{-i \omega_n\tau}  \sin n\sigma
\end{aligned}
\end{equation}
for an open string stretched between two D-strings located at the
transverse positions $x_0$ and $x_1$. 

Therefore, we see that the open strings attached to the D-string do
not have zero modes in the Dirichlet direction (defined, strictly
speaking, as the $\sigma$-independent part of their Fourier
expansion). However, this property does not mean that the D-string
lacks the zero modes which allow it to move in the transverse
directions. The latter correspond to the fields $x^r_{0/1}$ appearing
in the expansion~\eqn{e:x_mode_expansion}.

Using the standard Poisson/Dirac brackets between the
bosonic/fermionic coordinates, one can perform the canonical
quantization of the open string ending on a D-string; for the
quarter-BPS D-string, the commutation relations between the
oscillators are
\begin{equation}
\label{e:commutation}
\begin{aligned}
\forall\, m, n \neq 0, 
~~~\left[a^r_m, a^s_n\right] = {\rm sgn}(n)\delta_{m+n,0}\, \delta^{rs}
~~~{\rm and}~~~
\{\theta^{1a}_m, \theta^{1b}_n\} = \frac{1}{4}(\Gamma^+)^{ab}\,\delta_{m+n,0} 
\end{aligned}
\end{equation}
where we have arbitrarily chosen the Fourier modes $\theta^1_n$ as
the independent fermionic variables.  These commutation relations can
in principle be used as a starting point to construct the open string
spectrum but we will refrain from addressing that problem here.

\section{Discussion and open issues}

In the present letter we have shown how to obtain boundary conditions
for the covariant open Green-Schwarz string which correspond to
D-branes in an Hpp-wave. We have shown that this analysis reproduces
all D-branes known from previous probe brane and supergravity
computations. The advantage of our approach is that, since we
start from a covariant system, we are able to produce
directly the global supersymmetry invariances of the action after
light-cone gauge fixing.

We have also discussed the properties of the curious
quarter-supersymmetric D-string. This object preserves dynamical
supersymmetries (only) and may therefore be consistent with
open/closed string duality. It is interesting to observe that similar
states appear in other backgrounds as well, for instance the
Nappi-Witten background with two planes, and it would be interesting
to study their quantum consistency as well. In view of the original
motivation to study Hpp-waves, it is also important to understand the
corresponding states in the dual gauge theory. Progress on this issue
will be reported elsewhere.

As a side result, we have mentioned that there seem to be two
different ways to obtain the D-string from boundary conditions on the
open string (see equation~\eqn{e:d1string}). It is at present not
clear to us whether these boundary conditions correspond to the same
physical object.

Finally, the present analysis and evidence in favour of the presence
of a D-string is likely to be an argument that supports the existence
of a D-instanton in the Hpp-wave background. The present literature on
the gauge dual of D-instanton induced vertices in the Hpp-wave
background is not sufficient yet to rule out such an object, and
further progress along the lines of~\cite{Bergman:2002hv} would be
welcome.

\section*{Acknowledgements}
We thank Nadav Drukker, Matthias Gaberdiel, Shiraz Minwalla, Robert
Myers, Marika Taylor and especially Michael Green for discussions. 
The work in this paper was partially sponsored by EU contracts
HPRN-CT-2000-00122 and HPRN-CT-2000-00148.

\vfill
\appendix
\section{Appendix}
\subsection{Details of the Green-Schwarz string in the Hpp-wave}

This appendix contains the details of the covariant action of the
Green-Schwarz string in the Hpp-wave. This mostly follows
Metsaev~\cite{Metsaev:2001bj} though or notation is slightly
different. From an analysis of the kappa symmetry transformations in
superspace, one can see that the only boundary terms arise from the
variation of the Wess-Zumino part of the action (see
footnote~\ref{f:noboundary} below). This part of the action is given
by
\begin{equation}
\label{e:SWZ}
S_{WZ} = -2i \int_0^1{\rm d}t \int\!{\rm d}^2\sigma\,
    \epsilon^{ij} E^r_{it} \big(\Theta^T {\cal C} \Gamma_r E_{jt}\big)
\end{equation}
where the super-vielbeine are expanded as~\cite{Metsaev:2001bj}
\begin{equation}
\begin{aligned}
E   &= \frac{\sinh{\cal M}}{{\cal M}} {\cal D}\Theta\, ,\\[1ex]
E^r &= e^r - 2i\,\bar\Theta \Gamma^r \frac{\cosh{\cal M}-1}{{\cal M}^2}
{\cal D}\Theta\, .
\end{aligned}
\end{equation}
The matrix ${\cal M}^2$ is given by
\begin{equation}
{\cal M}^2 = \begin{pmatrix}
A & B \\ -B^* & - A^*
\end{pmatrix}\, ,
\end{equation}
where ($i,j$ run over $1,2,3,4$ only)
\begin{equation}
\label{e:ABdef}
\begin{aligned}
A^a{}_b &= \sum_{i=1}^5 (L_i\theta)^a (\bar\theta R_i)_b \\[1ex]
        &= -( I \Gamma^+ \Gamma^r \theta)^a(\bar\theta\Gamma^\mu)_b 
           -(\Gamma^{+i}\theta)^a(\bar\theta\Gamma^i I)_b
           +\tfrac{1}{2}(\Gamma^{ij}\theta)^a(\bar\theta\Gamma^+\Gamma^{ij} I)_b\\[1ex]
           &\qquad + \{1,2,3,4\} \rightarrow \{5,6,7,8\}\, ,\\[1ex]
B^a{}_b &= \sum_{i=1}^5 (L_i\theta)^a (\theta R_i)_b \\[1ex]
        &= -( I \Gamma^+ \Gamma^r \theta)^a(\theta\Gamma^\mu)_b 
           +(\Gamma^{+i}\theta)^a(\theta\Gamma^i I)_b
           -\tfrac{1}{2}(\Gamma^{ij}\theta)^a(\theta\Gamma^+\Gamma^{ij} I)_b\\[1ex]
           &\qquad + \{1,2,3,4\} \rightarrow \{5,6,7,8\}\, .
\end{aligned}
\end{equation}
The matrix $I$ appearing above (denoted $\Pi$ in
\cite{Metsaev:2001bj}) and the associated $J$ are given by
\begin{equation}
\label{e:IandJdef}
I = \Gamma^{1234}\, , \quad J = \Gamma^{5678}\, .
\end{equation}
The two real fermions are grouped according to
\begin{equation}
\Theta = \frac{1}{\sqrt{2}}\begin{pmatrix} \theta_1 + i \theta_2 \\ \theta_1 -i \theta_2 \end{pmatrix}
\,,\quad
\bar\Theta = \frac{1}{\sqrt{2}}\begin{pmatrix} \theta_1 - i \theta_2 \\
\theta_1 + i \theta_2 \end{pmatrix}^T {\cal C}\, ,
\end{equation}
Inserting these expressions, keeping all the terms which lead to
variations up to and including fourth order in $\Theta$ and taking
care of the $t$-integral, one obtains
\begin{equation}
\label{e:compact_action4}
\begin{aligned}
S_{WZ} = -i \int\!{\rm d}^2\sigma\,\epsilon^{ij}
     \Big[ e^r_i \big( \Theta^T{\cal C} \Gamma_r {\cal D}_j\Theta \big)
           -\tfrac{1}{2}i \big(\bar\Theta &\Gamma^r {\cal D}_i\Theta\big)
              \big(\Theta^T{\cal C} \Gamma_r {\cal D}_j\Theta\big)\\[1ex]
        &   +\tfrac{1}{12} e^r_i \big( \Theta^T{\cal C} \Gamma_r {\cal M}^2
     {\cal D}_j\Theta\big)
     \Big]\,,
\end{aligned}
\end{equation}
Note that, due to the $t$-integral which leads to a different
coefficient for every power of $\theta$, the first line above does not
factorise as $E_i{}^r (\ldots)$. To this order in the fermions, one
thus finds two parts: one which is obtained from the flat-space action
by replacing the normal derivative with a covariant derivative, and
one which arises from a ${\cal M}^2$ term in the expansion of
$E^a$. These are, in terms of $\theta_{1,2}$, given by
respectively
\begin{equation}
\label{e:S1WZ}
S^1_{WZ} = \int\!{\rm d}^2\sigma\, \Big[
-i \epsilon^{ij} e_i{}^r 
  \big( \bar\theta_1 \Gamma_r \hat D_j \theta_1 
        -\bar\theta^2 \Gamma_r \hat D_j \theta^2 \big)
+ \epsilon^{ij} \big( \bar\theta_1\Gamma^r \hat D_i \theta_1\big)
                \big( \bar\theta_2\Gamma_r \hat D_j
\theta_2\big)\Big]\, ,
\end{equation}
and
\begin{equation}
\label{e:firstMsquared}
S^2_{WZ} = -\frac{1}{12}\int\!{\rm d}^2\sigma\,
  \begin{aligned}[t] {} & \epsilon^{ij}\Big[ 
       \big( \bar\theta_1 \Gamma^r  I \Gamma^+\Gamma^s \theta_2 
            +\bar\theta_2 \Gamma^r I \Gamma^+\Gamma^s \theta_1 \big)
       \big( \bar\theta_1 \Gamma_s \partial_i\theta_1 
            +\bar\theta_2 \Gamma_s \partial_i\theta_2\big)\\[1ex]
   {} &+\big( \bar\theta_1 \Gamma^r\Gamma^{+m} \theta_1 
           - \bar\theta_2 \Gamma^r\Gamma^{+m}\theta_2 \big)
       \big( \bar\theta_1\Gamma_m I\partial_i\theta_2 
              -\bar\theta_2\Gamma_m I\partial_i\theta_1\big)\\[1ex]
   {} &- \tfrac{1}{2}\big( \bar\theta_1 \Gamma^r\Gamma^{mn}\theta_1
                          -\bar\theta_2 \Gamma^r\Gamma^{mn}\theta_2\big)
            \big( 
                \bar\theta_1 \Gamma^+\Gamma_{mn} I\partial_i\theta_2
               -\bar\theta_2 \Gamma^+\Gamma_{mn} I\partial_i\theta_1
            \big)\\[1ex]
   {} &+ \{1,2,3,4\} \rightarrow \{5,6,7,8\} \Big] \partial_j X^\mu
 e_{\mu r}\, .
\end{aligned}
\end{equation}
Here $m,n$ run over $1\ldots 4$ only. We do not have to consider here
the terms in $\hat D_j \theta$ that are proportional to $\partial_i
X^\mu$, as these will produce variations of order $\theta^6$.  In
these expressions the hatted covariant derivatives are given by
\begin{equation}
\hat D_i \theta_{1,2} = \partial_i \theta_{1,2} 
+ \frac{1}{4}\omega_{\mu rs} \Gamma^{rs} \theta_{1,2} \partial_i X^\mu
\pm \frac{1}{2}e_\mu{}^r  I \Gamma^+ \Gamma_r \theta_{2,1}
\partial_i X^\mu\, .
\end{equation}
There are additional four-fermi terms in the action, but these will
not contribute to the boundary terms of the kappa variation at second
and fourth order in the fermions.

The variation $\delta_\kappa X^\mu$ can be expressed in terms of 
$\delta_\kappa \theta_{1,2}$ through the defining relation in 
superspace,
\begin{equation}
\label{e:defkappa}
\delta_\kappa X^M E_M{}^r = 0\, \quad\Rightarrow 
\quad \delta_\kappa X^\mu = -i \bar\theta_1 \Gamma^\mu \delta_\kappa\theta_1 - i \bar\theta_2
\Gamma^\mu \delta_\kappa\theta_2 + {\cal O}(\theta^4)\, .
\end{equation}
Upon variation of the action all bulk terms vanish because the
background is on-shell, and the remaining boundary terms are given by
\footnote{\label{f:noboundary} From~\eqn{e:defkappa} one can now show
that the only boundary terms come from the Wess-Zumino part of the
action: in the kinetic terms, all variations of the fields inside
a derivative are of the form
\begin{equation}
\partial_i ( \delta_\kappa X^M ) E_M{}^r = - \delta_\kappa X^M \partial_i E_M{}^r\, ,
\end{equation}
there is therefore no need for partial integration, and no boundary
terms are generated.}.
\begin{equation}
\begin{aligned}
\delta_\kappa S^1_{WZ} = \int_{\partial \Sigma}
  & \Big[ i\, \delta_\kappa X^\mu \big(
       \bar\theta_1 \Gamma_\mu \hat D \theta_1 
      -\bar\theta_2 \Gamma_\mu \hat D \theta_2 \big)\\[1ex]
&- i\, {\rm d} X^\mu \big(\begin{aligned}[t]
  & \bar\theta_1 \Gamma_\mu \delta_\kappa\theta_1 
  + \tfrac{1}{4} \bar\theta_1\Gamma_\mu \Gamma^{rs} \omega_{\nu rs}
       \theta_1 \delta_\kappa X^\nu 
  + \tfrac{1}{2} \bar\theta_1 \Gamma_\mu  I \Gamma^+ \Gamma_\nu
       \theta_2 \delta_\kappa X^\nu\\[1ex]
  -& \bar\theta_2 \Gamma_\mu \delta_\kappa\theta_2 
  - \tfrac{1}{4} \bar\theta_2\Gamma_\mu \Gamma^{rs} \omega_{\nu rs}
       \theta_2 \delta_\kappa X^\nu
  + \tfrac{1}{2} \bar\theta_2 \Gamma_\mu  I \Gamma^+ \Gamma_\nu
       \theta_1 \delta_\kappa X^\nu \big) \end{aligned}\\[1ex]
&- \big( \bar\theta_1 \Gamma^r \delta_\kappa\theta_1\big)
   \big( \bar\theta_2 \Gamma_r \hat D \theta_2 \big)
+  \big( \bar\theta_2 \Gamma^r \delta_\kappa\theta_2\big)
   \big( \bar\theta_1 \Gamma_r \hat D \theta_1 \big)
\Big] \, .
\end{aligned}
\end{equation}

\subsection{Kappa symmetry boundary terms at higher orders in theta}

With the boundary conditions~\eqn{e:proj1} for generic branes
and~\eqn{e:d1string} for the D-string, we have found that
\begin{equation}
(\Theta^T {\cal C} \Gamma_r {\cal D}_j \Theta) = 0 \quad \text{if
$r\in N$}\, .
\end{equation}
This is very useful for higher order calculations: it means that in
the variation of the first term in~\eqn{e:compact_action4} we do not
have to consider variation of the $e_i^r$ factor at all. However,
there are still other variations.

Just as in~\eqn{e:firstMsquared}, the higher order terms in ${\cal
M}^2$ will reduce to products of \mbox{$\theta$-bilinears}, each factor of
which is a sum or difference of two terms obtained by interchange of
$\theta_1$ and $\theta_2$. We need an efficient way to figure out the
relative signs between these two terms. One finds the following
general expression:
\begin{equation}
\begin{aligned}
\Theta^T{\cal C} \Gamma_r \big({\cal M}^2\big)^n {\cal D}\Theta 
 &= \begin{aligned}[t] \sum_{R_i, L_i}  &
 \Big[ \big(\theta \Gamma_r L_n \theta)\mp_n (\bar\theta \Gamma_r L_n \bar\theta\big)\Big]
  \\
 {}& \times  \Big[ \big(\theta R_{n} L_{n-1} \bar\theta)\pm_n \mp_{n-1}
 (\bar\theta R_{n} L_{n-1} \theta\big)\Big]\times
\cdots \\[1ex]
 {}& \times  \Big[ \big(\theta R_1 D\bar\theta)\pm_1 (\bar\theta R_1
 D\theta\big)\Big]\, ,\end{aligned}\\[1ex]
\bar\Theta \Gamma_r \big({\cal M}^2\big)^n {\cal D}\Theta 
 &= \begin{aligned}[t] \sum_{R_i, L_i}  &
 \Big[ \big(\bar\theta \Gamma_r L_n \theta)\mp_n (\theta \Gamma_r L_n \bar\theta\big)\Big]
  \\
 {}& \times  \Big[ \big(\theta R_{n} L_{n-1} \bar\theta)\pm_n \mp_{n-1}
 (\bar\theta R_{n} L_{n-1} \theta\big)\Big]\times
\cdots \\[1ex]
 {}& \times  \Big[ \big(\theta R_1 D\bar\theta)\pm_1 (\bar\theta R_1
 D\theta\big)\Big]\, .\end{aligned}\\[1ex]
\end{aligned}
\end{equation}
The $L_i$ and $R_i$ are defined in~\eqn{e:ABdef}. The signs $\pm_n$
are the relative signs between the \mbox{$n$-th} term in $A$ and the \mbox{$n$-th}
term in $B$, i.e. $\pm_n = \{ +, -, -, -, -\}$. The
expression~\eqn{e:firstMsquared} is a special case of the general
formula above.

Using these expressions, one can easily write down the expression for
the WZ~part of the action,~\eqn{e:SWZ}. An analysis of these seems to
suggest that even at higher order in $\theta$, the resulting boundary
terms in the variation under kappa symmetry all vanish when the
D-brane boundary conditions derived in the main text are
satisfied. This is in particular true for the D-string with either one
of the boundary conditions in~\eqn{e:d1string}.

\subsection{Conventions and notation}

We follow the conventions of Metsaev~\cite{Metsaev:2001bj}. The
$X^\pm$ coordinates are defined as \mbox{$X^\pm=(X^0\pm
X^9)/\sqrt{2}$}. The metric for the Hpp-wave is
\begin{equation}
{\rm d}s^2 = 2\,{\rm d}X^+ {\rm d}X^- - S ({\rm d}X^+)^2 + ({\rm
d}X^i)^2\, .
\end{equation}
The vielbeine for the PP-wave are 
\begin{equation}
e^{+} = {\rm d}X^+ \, , \quad
e^{-} = {\rm d}X^- - \frac{S}{2} {\rm d}X^+ \, , \quad
e^{r} = \delta_\mu{}^r {\rm d}X^\mu \, , 
\end{equation}
with $\eta_{+-}=1$, from which one computes the only non-zero
component of the spin connection,
\begin{equation}
\label{e:spinconnPP}
\omega_{+r} = - \tfrac{1}{2}\partial_r S\, {\rm d}X^+\, .
\end{equation}
For spinors $\theta$, which are positive chirality Majorana-Weyl, we
use a 32-component notation everywhere. The chirality projector and
Dirac bar are
\begin{equation}
\Gamma := \Gamma^{0\cdots 9}\, ,\quad 
\bar\theta = \theta^T {\cal C}\, ,
\end{equation}
where ${\cal C}$ is the charge conjugation matrix satisfying ${\cal
C}^T=-{\cal C}$ and ${\cal C}^{-1} (\Gamma^{\mu})^T {\cal C} = -
\Gamma^\mu$ in ten dimensions.  Our index conventions are summarised
in the following table:
\begin{center}
\begin{tabular}{l|cc}
                 & coordinate basis                   & tangent basis \\
\hline
superspace 10+32 dim.       &  $M,N,P,\ldots$                    & $A,B,C,\ldots$ \\
bosonic  10 dim.        &  $\mu,\nu,\rho,\ldots$             & $r,s,t,\ldots$ \\
fermionic  32 dim.      &  $\alpha,\beta,\gamma,\ldots$      & $a,b,c,\ldots$ \\
bosonic  2 dim.         &  $i,j,k$ & 
\end{tabular}
\end{center}

\bibliographystyle{JHEP}
\bibliography{missing}
\end{document}